# CODE SWARM: A CODE GENERATION TOOL BASED ON THE AUTOMATIC DERIVATION OF TRANSFORMATION RULE SET

Hina Mahmood[1], Atif Aftab Jilani[2] and Abdul Rauf[3]

[1]Department of Computing and Software, McMaster University, Hamilton, ON, Canada
[2]FAST-National University of Computer and Emerging Sciences, Islamabad, Pakistan
[3]Knightec AB, Västerås, Sweden

*ABSTRACT*

*Automatic generation of software code from system design models remains an actively explored research area for the past several years. A number of tools are currently available to facilitate and automate the task of generating code from software models. To the best of our knowledge, existing software tools rely on an explicitly defined transformation rule set to perform the model-to-code transformation process. In this paper, we introduce a novel tool named **Code Swarm**, abbreviated as **CodS**, that automatically generates implementation code from system design models by utilizing a swarm-based approach. Specifically, CodS is capable of generating Java code from the class and state models of the software system by making use of the previously solved model-to-code transformation examples. Our tool enables the designers to specify behavioural actions in the input models using the Action Specification Language (ASL). We use an industrial case study of the Elevator Control System (ECS) to perform the experimental validation of our tool. Our results indicate that the code generated by CodS is correct and consistent with the input design models. CodS performs the process of automatic code generation without taking the explicit transformation rule set or languages metamodels' information as input, which distinguishes it from all the existing automatic code generation tools.*

## 1. INTRODUCTION

The advent of Model Driven Engineering (MDE) [1] brought a paradigm shift in the history of software engineering by changing it from a code-centric to a model-centric activity. The use of models in the development of software has a rich history [2]. Different types of software models can be developed, depending upon their intended usage. After the emergence of Unified Modeling Language (UML) [3] as the de facto industry standard for developing object-oriented systems [4, 5], the use of the UML class and state models is typically considered inevitable in the development of software. This is because these two models are representatives of both the system's static structure as well as its dynamic behaviour.

Manual transformation of software design models into code is a time-consuming and tedious task. The code written by hand may not be fully compliant with the design models due to the chances of human error, thus increasing the risk of a system malfunction [6].

Consequently, researchers have extensively worked on developing approaches and tools for facilitating and automating the task of generating code from various system models. The automation of the code generation process results in increased productivity, improved efficiency and reduced errors [7]. It also helps the software engineers in the on-time completion and delivery of their software systems [4].

In this paper, we introduce a tool named *Code Swarm*, abbreviated as *CodS*, that is developed to automatically generate Java code from the input class and state models of a software system. This tool is based on our framework presented in [8] and the automatic code generation approach proposed in [9]. CodS is capable of generating complete software implementation code only by using these two models. The dynamic behaviour of a software can be incorporated in these models by the use of Action Specification Language (ASL) [10]. ASL is an implementation-independent, precise and rich action language which is capable of specifying all the processing to be carried out by an object-oriented system. We selected ASL because a human reader can quickly scan through it because of its simplicity and readability [10]. Our tool, CodS, takes the class diagram of a software system to generate the structural code. The dynamic behaviour and method bodies are specified by CodS using the state models and ASL respectively.

We perform the experimental validation of our tool by applying it to generate code for the industrial case study of an Elevator Control System. Our results indicate that the code generated by CodS is simple, understandable and consistent with the input design models. We compare our tool with some existing research-based automatic code generation tools. The significantly distinct and distinguishable feature of our tool is that it performs the automatic code generation process in the absence of an explicit transformation rule set. Moreover, CodS always proposes a transformation for the input model element. This is rather impossible in the existing rule-based code generation tools, in which the absence of a rule for the transformation of a model element results in a failure to generate code corresponding to that model element.

The rest of this paper is organized as follows: Section 2 presents our motivation for the development of our tool, CodS. Section 3 explains the architecture of CodS. Section 4 describes the steps that CodS performs in order to automatically generate code from the input design models. Section 5 elaborates the application of CodS to the Elevator Control System (ECS) and discusses our obtained results. Section 6 presents a brief comparison of our tool CodS with some existing rule-based code generation tools. Finally, Section 7 concludes this paper.

## 2. MOTIVATION

A plethora of tools that currently exist for automatic code generation rely on making use of a transformation rule set for bridging the gap between software modeling and implementation languages. These transformation rules create a mapping between the source modeling and target programming languages. In industrial organizations, where a record of past transformations exists, it is a general observation that experts often find it easy to use existing transformations as examples to perform new transformations [11, 12], instead of investing resources in defining, expressing and maintaining a complete, consistent and non-redundant transformation rule set.

Presently, transformation rules need to be manually defined by the domain experts. To the best of our knowledge, there does not exist a tool that supports the automatic derivation of transformation rules from existing transformation examples. Practically, the development of a transformation rule set is a complex and demanding activity. Some transformations cannot be easily expressed as rules [11]. In some situations, rule induction can become impossible or difficult to achieve [12]. This is especially true in situations where little domain knowledge is available [13]. Transformation rules are not only difficult to define, but also hard to express and maintain. With the passage of time, transformation rule sets may evolve. Adding new rules or changing existing rules make it complex to ensure their consistency and correctness [11].

The transformation rule set must be complete, consistent, non-redundant and correct in order to ensure the generation of accurate software implementation code. The absence of a rule for transforming a single model element from the transformation rule set can hinder the transformation process by resulting in a failure to generate code corresponding to that model element. From all these observations, our work starts where we view the automatic code generation problem as the one that could be solved with *fragmentary knowledge*, i.e. with existing examples of model-to-code transformations. In this case, there should exist a tool that can automatically derive transformation rules from the existing set of transformation examples, rather than compelling the organizations to base their transformations on an explicit user-defined rule set.

In order to improve this situation, we develop a tool named *CodS* in Java. Our tool should allow industrial organizations to utilize the existing set of transformation examples, accumulated from their past experiences, to perform the current code generation process. Instead of relying on the explicit user-defined transformation rule set, CodS uses these examples to automatically derive transformation rules from them, and generate code for the current input models. Specifically, the existing set of transformation examples are used to train the system regarding automatic code generation. After the system is trained, the input models, which are to be transformed, are provided as input and the corresponding implementation code is automatically generated by CodS without experts' intervention.

## 3. CodS Architecture

Figure 1 gives an overview of the architecture of our tool, CodS, which is based on our framework presented in [8] and our automatic code generation approach described in [9]. The architecture of CodS has three major components: *PredicateParser*, *SearchEngine* and *TransformationEngine*. CodS takes a set of existing transformation examples and the source models to be transformed as input. This input is managed and organized by the PredicateParser. The SearchEngine component finds an optimal transformation solution for the input models by using the available transformation examples. The generated optimal solution is then used by the TransformationEngine to transform the input model constructs into target code predicates. Finally, these code predicates are transformed into complete Java code statements. In this way, a set of files containing the code predicates and the complete Java code is produced as the final output by CodS.

### 3.1. PredicateParser

The component of PredicateParser initiates the execution of CodS. It takes a set of transformation examples and the models to be transformed as input. The input transformation examples are represented as mapping blocks and used to build a knowledge base. A *mapping block* relates the subset of source model constructs to their equivalent constructs in the target implementation language. The component of the PredicateParser has two primary responsibilities: 1) to divide the mapping block predicates into source model constructs and target code constructs, and 2) to organize and store this knowledge into separate structures.

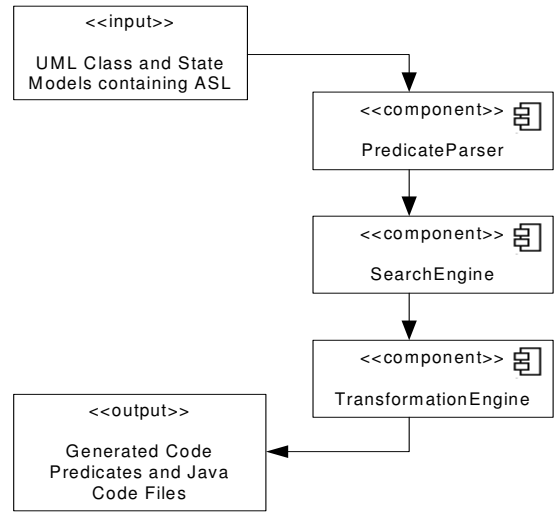

Figure 1. CodS architecture

A set of class and state models containing ASL statements is also taken as an input by the PredicateParser. Just like the transformation examples, these source models are expressed as predicates as well. These input model predicates are also divided into separate model constructs by the PredicateParser.

## 3.2. SearchEngine

The SearchEngine is the major and most important component of CodS. Based on our swarm-based automatic code generation approach presented in [9], we use the *Particle Swarm Optimization (PSO) algorithm* [14] as the search engine of CodS. The core responsibility of the SearchEngine is to search for an optimal transformation solution for the input source models. It uses the source model constructs, structured by the PredicateParser, to search for their matching transformations in the knowledge base. The transformation solution is in the form of a set of mapping block numbers, one mapping block corresponding to each input model construct.

The SearchEngine initializes by assigning random mapping blocks for transforming the source model constructs. The quality of this random solution is assessed by using the fitness-function of PSO defined in [9]. Depending upon the fitness value calculated by the objective function, the parameters of PSO are updated and more solutions are generated in the next iterations. The SearchEngine component remains active until the total number of iterations of PSO is complete. In the end, the solution having the maximum fitness value is selected as the final optimal solution. In CodS, the optimal solution is represented as an array, with mapping block numbers as the array elements. The size of the array is equal to the total number of source model constructs.

## 3.3. TransformationEngine

Just like the SearchEngine, the TransformationEngine is also a significantly important component of CodS. The principal job of the TransformationEngine is to produce the target code, both in terms of predicates and code statements, corresponding to the input source models. The optimal solution generated by the SearchEngine is taken as an input by the TransformationEngine. For each input model construct, it searches for the matching construct in the mapping block selected in the optimal solution. The code predicate of this mapping block construct is then used to transform the input model construct, for which the mapping block is selected. In this way, the output produced by the TransformationEngine is a set of code predicates produced for the input source models.

This component of CodS is also responsible for transforming these code predicates into complete Java statements. The automatic conversion of the code predicates into code statements eradicates the manual effort and time required for the conversion. The complete Java code generated by the TransformationEngine is organized and stored as a set of Java files.

## 4. THE CODE GENERATION PROCESS

The CodS tool manages the activity of automatic code generation in terms of separate projects. Before initiating the code generation process, the designers first need to create a new CodS project from the CodS main interface. After that, they can perform the process of automatic code generation by following the four key steps. We describe each of these steps below.

### 4.1. Import Training Data

After creating a new CodS project, designers first need to import the training data to build a knowledge base. This is done by using the option available in the menu bar and toolbar of the CodS main interface. Training data may comprise of multiple text files containing a set of mapping blocks. When users browse the training data files, these files are automatically made a part of the current project by storing them in the 'Input' folder of the CodS current project.

### 4.2. Import Input Models

The second step in the code generation process is to import the input source models which are to be transformed into the target code. Currently, CodS requires that all the input source models must be present in a single text file, i.e. all class and state models along with the ASL statements belonging to one software system must be stored in one text file. While importing the input models' text file, this file is also stored automatically in the 'Input' folder of the active CodS project.

### 4.3. Transform Models

After providing input to CodS, designers need to select the option for transforming the input models. This is the most important step among all the four steps of the automatic code generation process performed by CodS. In this step, CodS gets trained to automatically derive rules from the available set of transformation examples. After the training, an optimal solution is generated by the use of the PSO algorithm. This optimal solution is then used to transform the input model constructs into target code predicates. These code predicates are stored in a text file named 'Predicates'. This text file is stored in the 'Output\Code Predicates' folder of the CodS current project.

At this step, a 'readme' text file is also generated by CodS. This readme file contains information related to the transformation process for later reference by the designers. Particularly, it contains the following information: 1) the number of input model constructs, 2) the total number of mapping blocks in the training data, 3) the best fitness value for the generated optimal solution, 4) the mapping block numbers selected in the optimal solution corresponding to each input model construct, 5) the number of evaluations performed by the PSO algorithm, and 6) the details of the source model construct and the mapping block, if an exact match of the input model construct is not found in the selected mapping block. This information is also displayed to users on the console of the CodS tool.

### 4.4. Generate Code

The final step of the automatic code generation process is performed by CodS when designers select the option of generating Java code corresponding to the code predicates that were produced in the previous step. The output produced at this step is a set of code files containing the complete Java code statements. These code files are stored in the 'Output\Java Code' folder of the currently opened CodS project. Moreover, another 'readme' text file is generated at this step. This readme file contains information about the input model constructs for which partial or no exact match was found in the mapping blocks, and their corresponding Java file names.

## 5. EXPERIMENTAL VALIDATION

This section describes our experimental setting for the validation of our tool, CodS, and discusses the obtained results.

### 5.1. Elevator Control System

We used the industrial case study of an Elevator Control System (ECS) to perform the experimental validation of our tool, CodS. The class model of the ECS, shown in Figure 2, consists of 8 classes and 4 sub-classes. The changing behaviour of the ECS in response to the user interaction is represented in the ECS state model [9]. This model consists of 13 states and 27 transitions. The actions performed within each state are specified using ASL. Currently, the class and state models of the ECS need to be manually expressed as predicates. We use the class and state models of 9 other software systems as the training data. These models are represented as 62 mapping blocks stored in 9 text files. Please refer to [9] for complete details.

### 5.2. Results and Discussion

By following the automatic code generation process described in Section 4, we generated code predicates and Java code files for the ECS using our tool, CodS. Figures 3-4 show the screenshots of CodS after performing the steps of model transformation (step 3) and code generation (step 4) respectively. In Figure 4, a list of all the generated Java classes is shown in the left panel of the CodS interface. These code files can be viewed by users in the center panel of CodS. The results for the automatic code generation process of the ECS performed by CodS are summarized in Table 1.

Table 1. Automatic code generation results for the Elevator Control System

| Mapping blocks | Model constructs | Correctly transformed constructs | Best fitness |
|---|---|---|---|
| 62 | 364 | 360 | 0.9780 |

Figure 2. Class diagram for the Elevator Control System (ECS)

Figure 3. CodS interface: Transform models

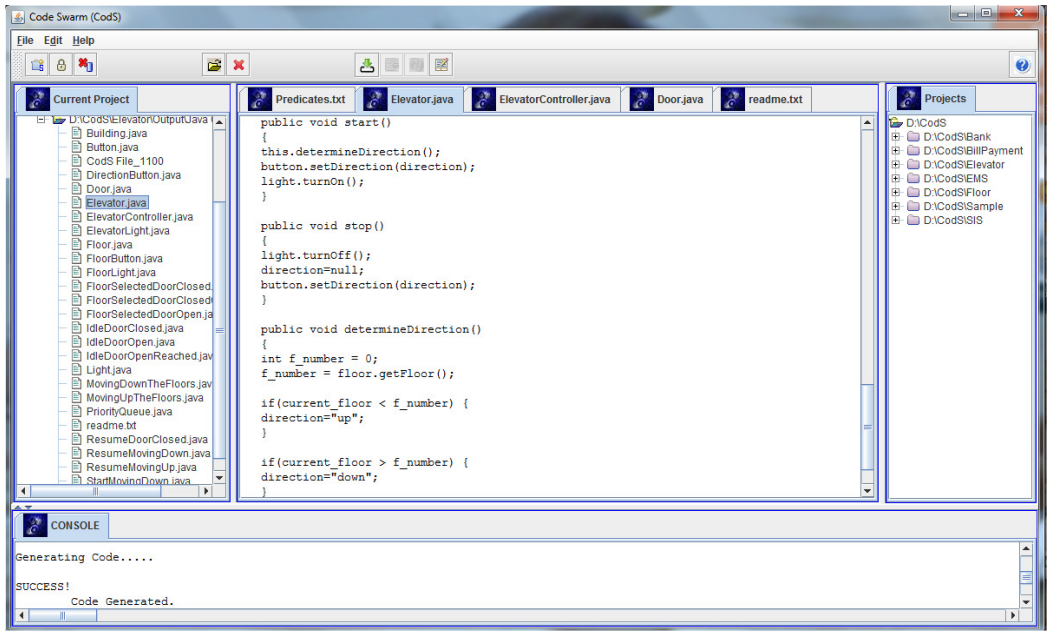

Figure 4. CodS interface: Generate code

By looking at the details of the obtained results, we note that out of 364 total input model constructs that needed to be transformed, there are 8 constructs for which no exact transformation was present in the training data. CodS highlights these 8 constructs on the console of its main interface and also stores this information in the readme text file to be referenced later by the designers, whenever required. Moreover, our tool is intelligent enough to look for the nearest matching transformations, in case if no exact transformation is found in the training data. Therefore, it suggests transformations for 5 of these 8 constructs. Of these 5 suggested transformations, 4 are found to be correct. In this way, the total number of correctly transformed constructs raises to 360.

The remaining 3 input model constructs for which no nearest match was found in the training data, they are mentioned separately by CodS for the designers' attention. Thus, complete details of the entire code generation process are made visible to the designers by CodS. This is beneficial, as designers can take further actions for the model constructs for which no corresponding code could be generated. In this way, the generated code remains consistent with the input models and designers are also made aware of the input model constructs that have been transformed correctly and the constructs for which no exact match was found in the given training data.

## 6. COMPARISON

In this section, we compare our tool CodS with the existing research-based automatic code generation tools. These tools include UJECTOR [4], JCode [15], dCode [16], OCode [17] and AutoKode [18]. The parameters and details of comparison are given below.

### 6.1. Underlying Approach

All the five comparison tools under discussion are based on the approaches that are input model-specific. However, CodS uses a generic automatic code generation approach presented in [9]. In other words, our approach can be used to transform any set of source models into any target programming language, provided that the transformation examples exist.

### 6.2. Behavioural Action Specification

In UJECTOR, actions need to be specified in the UML superstructure. The use of UML superstructure actions raises the level of complexity, as it is difficult and time-consuming to specify and understand these actions. AutoKode uses additional models like UML activity and sequence diagrams for modeling the system's dynamic behaviour and generating class and method definitions. On the other hand, CodS relies on a light-weight action language ASL for specifying the behavioural actions, which is simple, readable, and easy to learn and comprehend [10].

### 6.3. Explicit Transformation Rules

UJECTOR, JCode, dCode, OCode and AutoKode generate the implementation code by creating a mapping between the source modeling and the target programming languages. All these tools rely on the explicit specification of the transformation rule set. However, our tool CodS does not take a set of transformation rules as input. Rather, it is intelligent enough to automatically derive transformation rules from the existing set of transformation examples. Besides the training data, no extra information is needed by CodS to perform the automatic code generation process.

### 6.4. Exhaustive Rule Set

All five tools under discussion utilize an exhaustive set of transformation rules to correctly generate the target code. These tools fail to perform the transformation if a rule does not exist for any source model construct. However, CodS is smart enough to assist the designers by proposing a nearest transformation strategy, if no exact transformation match is found in the training data for any given input model construct.

## 7. CONCLUSIONS AND FUTURE WORK

This paper describes a novel tool named **Code Swarm**, abbreviated as **CodS**, which is based on our framework described in [8] and our swarm-based approach for automatic code generation presented in [9]. Our tool, CodS, is capable of generating complete Java code from the class models and state models that encapsulate actions using the Action Specification Language (ASL). The class model is used to generate the skeleton code, whereas state models and ASL action statements are used to specify the system's dynamic behaviour. An industrial case study of the Elevator Control System is used to perform the experimental validation of CodS. Our results indicate that the code generated by CodS is correct and consistent with the input design models. We also compare CodS with some existing research-based code generation tools. The most distinguishing feature of CodS is that it does not rely on using an explicitly defined transformation rule set for performing the model-to-code transformation process. Rather, it is intelligent enough to automatically derive rules from the existing set of transformation examples to perform the automatic code generation process.

Currently, CodS is capable of generating Java code from the input design models. In future, we intend to enhance the functionality and utility of CodS by adding support for more programming languages such as C++, etc. Also, presently, designers need to manually express their input class and state models which are to be transformed and the training data as predicates. We plan to automate this task in future to further facilitate the software designers and CodS users.

**AUTHORS**

**Hina Mahmood** has recently completed her Ph.D. in software engineering at McMaster University, Canada. She has served as the member program committee for several international conferences and journals. Her research interests include model-driven software engineering, supervisory control of discrete-event systems, model transformation, and formal modeling and verification of software control systems.

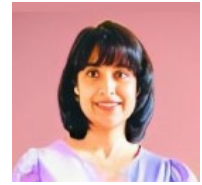

**Atif Aftab Jilani** is an Assistant Professor in the Department of Software Engineering at the National University of Computer & Emerging Sciences (Fast-NU), Islamabad, Pakistan. His research interests include model-driven systems and software engineering, search-based software testing, automated software testing, the application of ML/AI strategies in software engineering, and empirical software engineering.

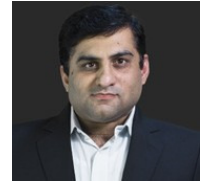

**Abdul Rauf** is a seasoned software engineering researcher and practitioner, boasting more than four years of hands-on expertise in applied industrial research and development. Additionally, he brings over a decade of experience in academia at the university level across international settings. He is currently working as Consultant with Knightec AB, Sweden.

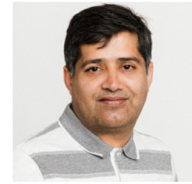